\documentclass[preprint,aps,prd,nofootinbib,superscriptaddress,tightenlines]{revtex4}

\usepackage{bm}
\usepackage{amsmath}
\usepackage{float}
\usepackage{dcolumn}
\usepackage{amssymb}
\usepackage{graphics}
\usepackage{amsmath, amsfonts, amsthm, amssymb, graphicx,comment}
\usepackage{etex}
\usepackage{hyperref,subfigure,color}
\PassOptionsToPackage{unicode}{hyperref}
\PassOptionsToPackage{naturalnames}{hyperref}

\begin{document}

\title{Perturbations of black holes in mass-varying massive gravity}
\author{De-Jun Wu}
\email[]{wudejun10@mails.ucas.ac.cn}
\affiliation{School of Science, Inner Mongolia University of Science and Technology, Inner Mongolia, Baotou 014010, China}

\begin{abstract}
Based on eigendecomposition and matrix perturbation theory, we propose an alternative method for analyzing perturbations in massive gravity that is applicable to both the base theory and its various modifications. We demonstrate the validity of this approach even in seemingly singular regimes. Applying this framework to mass-varying massive gravity, we compute perturbations for all constant-scalar  Schwarzschild-(A)dS black hole solutions. We show that, under the condition $\beta = \alpha^2$, the tensor sector of these black holes reduces to that of general relativity, while the scalar sector remains free from tachyonic instabilities given appropriate parameter choices. A deeper relationship is also revealed between the solutions with generic parameters and those with $\beta = \alpha^2$: their solution manifolds intersect at a specific locus where the perturbation vanishes.
\end{abstract}
\maketitle
\section{Introduction}

Driven by the direct detection of gravitational waves \cite{LIGOScientific:2016aoc}, black hole observations now enable precise  tests of general relativity \cite{Cardoso:2025npr} and place constraints on modified theories of gravity, including those with a non-zero graviton mass \cite{LIGOScientific:2016lio, LIGOScientific:2021sio},  offering a valuable empirical window into strong-field gravity and potential deviations from standard theoretical frameworks \cite{Yunes:2024lzm}.  To date, the only known class of self-consistent theories describing a massive graviton is based on the de Rham–Gabadadze–Tolley (dRGT) massive gravity framework introduced and developed in \cite{deRham:2010ik, deRham:2010kj,Hassan:2011hr,Hassan:2011ea}.  Extensive treatments of this area are provided in \cite{Hinterbichler:2011tt, deRham:2014zqa}. 

Black holes in massive gravity and its extensions have been extensively studied, 
including their perturbations, with much of the literature focusing on bi-gravity and multi-gravity. Comprehensive reviews can be found in \cite{Babichev:2015xha, Wood:2024acv}. 
A central challenge in analyzing black hole perturbations within massive gravity lies in computing the perturbation of    $\mathcal{K}^\mu_\nu$, which contains a matrix square root.  Over the years, several methods have been developed to address this issue \cite{Guarato:2013gba, Kodama:2013rea,  Bernard:2014bfa, Bernard:2015mkk, Kobayashi:2015yda, Cusin:2015tmf, Bernard:2015uic, Mazuet:2017hey, Mazuet:2018ysa, Wood:2025ujj}. One direct approach involves analytically obtaining the Fréchet derivative of the matrix square root via the Sylvester equation. Alternatively, the vielbein formalism can circumvent the complications associated with the matrix square root, although it requires sophisticated schemes to track physical degrees of freedom.  
However,  applying the existing methods to perturbation analysis becomes prohibitively complicated for  black hole solutions in mass-varying massive gravity (MVMG),  in which the fixed graviton mass term is replaced by a scalar potential, enabling more viable cosmological scenarios \cite{Huang:2012pe}. 
The solutions  considered in this work  \cite{Tolley:2015ywa, Zhang:2017jze} are similar to those discussed in \cite{Koyama:2011yg, Berezhiani:2011mt, Kodama:2013rea} but differ computationally from bi-gravity solutions in a significant respect, due to differences in the choice of reference metric:  $\mathcal{K}^\mu_\nu$ assumes a  more intricate form in MVMG, whereas it remains relatively simple in bi-gravity. 
To overcome this difficulty, we develop an alternative method based on eigendecomposition. As will become clear in subsequent sections, this approach yields notably simpler expressions, particularly for Schwarzschild-(A)dS backgrounds.   Moreover,   since the graviton potential depends solely on the spectrum of   $\mathcal{K}^\mu_\nu$,  this approach offers distinct advantages for studying the stability of hairy black holes in MVMG. It provides a unified framework for calculating the graviton potential, the effective energy-momentum tensor, and their perturbations—all of which are essential for such analyses. 

A limitation of this method is that it requires $\mathcal{K}^\mu_\nu$ to be diagonalizable. We attempted to relax this restriction by generalizing the eigendecomposition to a Schur decomposition in perturbation calculation but were unsuccessful. Furthermore, the method is highly sensitive to eigenvalue degeneracies of $\mathcal{K}^\mu_\nu$, which may vary across a continuous family of solutions. In such cases, the perturbations appear to exhibit singular behavior. However, we demonstrate that these transitions are in fact smooth, at least to leading order. 

We apply this method to investigate  the stability of  trivial (i.e., constant-scalar)  Schwarzschild-(A)dS solutions  in MVMG. We find that for $\beta = \alpha^2$, these black holes are  stable, given a reasonable choice of scalar potentials.  As an initial step toward understanding the stability of hairy black holes in MVMG, this study lays the groundwork for subsequent research into phenomena such as spontaneous scalarization and superradiance in MVMG.  
 
The paper is organized as follows. In Sec.~\ref{setup}, we briefly review the MVMG model  and  background solutions.
In Sec.~\ref{ptb_m}, we formulate the perturbation equations in terms of eigendecomposition. We apply this method to study black hole stability in Sec.~\ref{stability} and present our conclusions in Sec.~\ref{conclusion}. Detailed calculations and supplementary discussions are provided in the appendices.

\section{setup} \label{setup}
For mass-varying massive gravity, 
\begin{equation}
    S = \frac{{M_P^2 }}{2} \int {d^4 x} \sqrt { - g} \left[ {R + V\left( \varphi  \right) U({\cal K}) - \frac{1}{2}g^{\mu\nu} \partial _\mu  \varphi \partial_\nu   \varphi  - W\left( \varphi  \right)} \right],
\end{equation}
where $U({\cal K})\equiv U_2  + \alpha _3 U_3  + \alpha _4 U_4$, and
\begin{align*}
    U_2  &= {\cal K}_{[\mu }^\mu  {\cal K}_{\nu ]}^\nu,\\
    U_3  &= {\cal K}_{[\mu }^\mu  {\cal K}_\nu ^\nu  {\cal K}_{\rho ]}^\rho,\\
    U_4  &= {\cal K}_{[\mu }^\mu  {\cal K}_\nu ^\nu  {\cal K}_\rho ^\rho  {\cal K}_{\sigma ]}^\sigma,
\end{align*}
with ${\cal K}^\mu_\nu$ defined as ${\cal K}_\nu ^\mu   = \delta _\nu ^\mu   - \sqrt {g^{\mu \rho }  \eta_{\rho\nu}}$. $\eta_{\rho\nu}$ is the reference Minkowski metric. The equations of motion are
\begin{align}
    \label{eisteq}
G_{\mu\nu} &=\frac{1}{2}T_{\mu\nu}+V X_{\mu\nu},
\\
 \label{eompsi}
\frac{1}{{\sqrt { -g } } }\partial _\mu \left( {\sqrt {-g } g^{\mu\nu}  \partial_\nu \varphi } \right) &= W_\varphi-V_\varphi U,
\end{align}
where $W_\varphi=\partial W/\partial\varphi$, $V_\varphi=\partial V/\partial\varphi$, $T_{\mu\nu}$ is  the energy-momentum tensor of the scalar field.
\begin{equation}
    T_{\mu\nu} = \partial_\mu \varphi \partial _\nu \varphi  - g_{\mu\nu}\left(  \frac{1}{2}g^{\rho\sigma} \partial _\rho  \varphi \partial_\sigma   \varphi + W \right),
\end{equation}
and the effective energy-momentum tensor is
\begin{align}
     X_{\mu\nu}  &= \frac{1}{2}\bigg[-({\cal K}_{\mu\nu}  - [{\cal K}] g_{\mu\nu})  + \alpha \left( {{\cal K}_{\mu\nu}^2  - [{\cal K}] {\cal K}_{\mu\nu}  + U_2 g_{\mu\nu} } \right) \nonumber\\
  &- \beta \left( {{\cal K}_{\mu\nu}^3  - [{\cal K}] {\cal K}_{\mu\nu}^2  + U_2 {\cal K}_{\mu\nu}  - U_3 g_{\mu\nu} } \right)\bigg],
\end{align}
 in which two new parameters are defined as $\alpha=1+\alpha_3$,  $\beta=\alpha_3+\alpha_4 $. 

We introduce a static, spherically symmetric ansatz.
\begin{align}
\label{metric}
d s^2  &=  - a(r)dt^2 + 2b(r)drdt  + e(r)dr^2   + f(r) r^2 d\Omega ^2,
\\
\label{metric2}
d s^2_\eta &= \eta_{\mu\nu}dx^{\mu}dx^{\nu}  =  - dt^2  + dr^2   +r^2 d\Omega ^2,
\\
\label{varphian}
\varphi &= \varphi(r).
\end{align}
Several methods exist to address the matrix square roots arising in the field equations. We adopt an approach that computes the matrix square root via eigendecomposition. This method is  advantageous in the sense  that it provides a unified treatment for both the background and perturbation equations. 
 
The eigenvalues of ${\cal K}^{\mu}_{\nu}$ are 
\begin{align}
\label{k1}
    k_1  &= 1 - \sqrt{ \frac{2}{ {z + x }  } },
\\
\label{k2}
k_2 &= 1 - \sqrt{ \frac{2}{ {z -x }  } },
\\
\label{k3k4eq}
k_3&=k_4=1 - \sqrt {\frac{1}{f}},
\end{align}
where we have defined $x=\sqrt {z^2  - 4c} $  and  $z = a+e$,     $c= a e+b^2$.     The definition of ${\cal K}^{\mu}_{\nu}$ ensures that $k_i$ is uniquely determined.  The corresponding eigenvectors are 
\begin{equation}
\label{eigenvt}
    \begin{pmatrix} 1\\bn\\0\\0 \end{pmatrix},
\begin{pmatrix} bn\\1\\0\\0 \end{pmatrix},
\begin{pmatrix} 0\\0\\1\\0 \end{pmatrix},
\begin{pmatrix} 0\\0\\0\\1 \end{pmatrix},  
\end{equation}
with $n= \frac{2}{x+y}\neq 0$ and $y=a-e$. The choice of eigenvectors is not unique; here, we present only the set with the simplest form. For computing $X^\mu_\nu$, this choice is immaterial, as all valid sets yield identical results. However, calculating the perturbation of $X^\mu_\nu$ requires a specific class of eigenbasises, which we will adopt when necessary. 
 
In matrix form, the effective energy-momentum tensor is expressed as  
\begin{equation}
\label{X}
    X = P  \bar X  {P^{ - 1}},
\end{equation}
where $X$ denotes the mixed-index tensor $X^\mu_\nu$, and $\bar X=\mathrm{diag}(k_1, k_2, k_3, k_4)$ is its diagonalized form with eigenvalues $k_i$.
The matrix $P$ is invertible, with its columns formed by the eigenvectors of ${\cal K}^{\mu}_{\nu}$.

Hairy solutions exist in which the graviton mass is large near the black hole and decays with increasing radial distance \cite{Tolley:2015ywa, Zhang:2017jze}. Alternatively, setting $\varphi(r)=0$ yields  trivial solutions, provided that this configuration corresponds to a local minimum of both $V$ and $W$. This leads to two distinct classes of Schwarzschild-(A)dS solutions. The first class arises only under the specific parameter choice $\beta = \alpha ^2$ and can be expressed as Schwarzschild-(A)dS solutions in either the Gullstrand–Painlevé or Eddington–Finkelstein frame \cite{Berezhiani:2011mt}. The second class exists for generic parameter values and admits an explicit analytical form \cite{Koyama:2011yg}. Crucially, these two classes exhibit fundamentally different perturbative behaviors. A detailed review of these background solutions is provided in Appendix~\ref{bgsol}.
 
\section{perturbation} \label{ptb_m}

In this section we develop a novel method to compute $\delta X^\mu_{\nu} $.  To this end, we perturb Eq.~\eqref{X} to obtain  
\begin{align}
\label{perX}
    \delta X &= \delta P'    \bar X   {P'^{ - 1}} + P'   \delta \bar X   {P'^{ - 1}} + P'   \bar X   \delta {P'^{ - 1}} \nonumber\\ 
    & = \delta P'   {P'^{ - 1}}   X + P'   \delta \bar X   {P'^{ - 1}} + X   P'   \delta {P'^{ - 1}}. \nonumber\\
     &= P'   \delta \bar X   {P'^{ - 1}}+\left[\delta P'   {P'^{ - 1}}, X  \right].
\end{align}
Square brackets denote the matrix commutator. The matrix   $  P'$ comprises the eigenvectors appropriate for the perturbation analysis, specifically those associated with degenerate eigenvalues; a detailed discussion follows below. 

$ \delta \bar X$ is a diagonal matrix whose entries consist of $ \delta k_i $.  $ \delta k_i $ is related to the perturbation of the eigenvalues of matrix $ g^{\mu \rho }  \eta_{\rho\nu}$, denoted as $\lambda_i$, by the equation $\delta k_i=-\frac{1}{2 \sqrt{\lambda_i}} \delta \lambda_i= \frac{1}{2 (k_i-1)} \delta \lambda_i$. This relation holds for both distinct and degenerate eigenvalues, provided that  $\lambda_i>0$.  It remains to establish the relationship between $\delta  \lambda_i $ and the metric perturbation. 

The computation of eigenvalue perturbations induced by small matrix variations is analogous to perturbation theory in quantum mechanics and has been extensively studied in matrix perturbation theory.  For a simple (non-degenerate) eigenvalue,  the first-order correction admits a closed-form expression.   Let $\lambda$ be a simple eigenvalue of the  matrix $g^{-1 }  \eta$ with corresponding right eigenvector $v$ and left eigenvector $u$ which is a row vector. Then 
\begin{align}
     \delta \lambda  = \frac{{u\delta g^{-1 }  \eta v}}{{uv}}.
\end{align}
For multiple eigenvalues, one may adopt an approach analogous to degenerate perturbation theory.  Upon introducing the perturbation  $\delta g^{-1}  \eta$,  it is first projected onto this eigenspace. Following the formalism of quantum mechanics, we construct a matrix $\tilde \Omega $ with elements 
$ \tilde \Omega_{ij}= u_i \delta g^{-1 }  \eta  v_j$,   where $u_i$ and $v_j$ are left and right eigenvectors,   corresponding to the degenerate eigenvalue.  To simplify subsequent expressions, the condition $ u_i  v_j  = \delta_{ij} $ is  imposed; however, explicit normalization of the individual vectors $v_i$ is not necessary.   Provided that $\mathcal{K}$ is diagonalizable, such left eigenvectors always exist and can be identified directly from the rows  of $P^{-1}$. 
The eigenvalues of this projected matrix correspond to the first-order perturbations of the original degenerate eigenvalues.  Specifically, each 
 $\delta \lambda_i$  equals  an eigenvalue of $\tilde \Omega$. 

To prevent discontinuities in the eigenvectors that would invalidate the perturbation expansion, the unperturbed basis must be chosen such that   $\tilde \Omega $  is diagonal.  
Because the unperturbed matrix $g^{-1}  \eta$ acts as the identity operator on the eigenspace associated with the degenerate eigenvalue, it naturally shares eigenvectors with $\tilde \Omega$.
We denote this adapted set of eigenvectors by   $P'$  in Eq.~\eqref{perX}. 
It is given by   $ P' = P A$, where $A$ is a linear transformation  matrix whose columns are the eigenvectors of   $\tilde \Omega $.    Strictly speaking, this method  requires  $\tilde \Omega $ to  be diagonalizable.  In practice, however, computations typically involve only the trace of  $\tilde \Omega $ and the matrix itself.  Therefore, it is reasonable to assume the results obtained via this approach remain generally applicable even when $\tilde \Omega $ is non-diagonalizable.

We remark that $\delta P'   {P'^{ - 1}}$ possesses the structure of a connection form. While this geometric perspective suggests more elegant formulations via differential geometry, such an analysis lies beyond the scope of the present work. For our purposes, it suffices to note that its matrix elements admit a simple and explicit expression. Assuming  the perturbation of an eigenvector does not contain contributions from itself,  the element of $\delta P'   {P'^{ - 1}}$ can be written as 
\begin{align}
    {\left( {\delta P'{P'^{ - 1}}} \right)_{\alpha \beta }} = \sum\limits_{i \ne j} {{{{\bf{v'}}}_{\alpha i}}} \frac{{{{u'}_i}   \delta {g^{ - 1}}\eta    {{v'}_j}}}{{{\lambda _j} - {\lambda _i}}}{{\bf{u'}}_{j\beta }},
\end{align}
where ${\bf{v'}}_{\alpha i} $ is the $\alpha$-th component of the $ v_i$, which is also the $(\alpha,i) $ entry of matrix $P'$. Similarly  ${{\bf{u'}}_{j\beta }}$ is the corresponding entry of matrix $P'^{ - 1}$.  If we define a projection as
\begin{align}
  ( {\rm P'}_i)_{\alpha \beta} = {\bf{v'}}_{\alpha i}  {\bf{u'}}_{i \beta},
\end{align}
this expression can be further simplified to:
\begin{align}
\label{dpp}
     {\delta P'{P'^{ - 1}}}  = \sum\limits_{i \ne j} \frac{{\rm P'}_i   \delta {g^{ - 1}}\eta    {\rm P'}_j}{{\lambda _j} - {\lambda _i}}.
\end{align}
For a non-degenerate eigenvalue, the projection remains unchanged under perturbation, i.e.,  ${\rm P'}_i = {\rm P}_i$.  For projections associated with a degenerate eigenvalue, it is more convenient to work with their sum:    
\begin{align}
 \sum\limits_{i} \frac{{\rm P'}_i   \delta {g^{ - 1}}\eta    {\rm P}_k}{{\lambda _k} - {\lambda _i}}
  &= \frac{1}{{\lambda _k} - {\lambda^* }}\left( \sum\limits_{i } {\rm P'}_i \right)  \delta {g^{ - 1}}\eta    {\rm P}_k,
\end{align}
where we used the fact that the projections correspond to the same eigenvalue, which we denote as $\lambda^*$.  The perturbed eigenvectors are linear combinations of the unperturbed ones within the degenerate subspace, expressed as $ {\bf{v'}}_{\alpha i}=  \sum\limits_{ j}
{\bf{v}}_{\alpha j}A_{ji}$, where $A_{ji}$ are the mixing coefficients.  Correspondingly, the left eigenvectors transform as  $  {\bf{u'}}_{i \beta} = \sum\limits_{j} A^{-1}_{ij} {\bf{u}}_{j \beta}$.  It follows directly that  $\sum\limits_{i } {\rm P'}_i = \sum\limits_{i } {\rm P}_i $, since this sum is precisely the projector onto the degenerate eigenspace, which is invariant under any change of basis within that subspace.   Consequently, all projections based on the new eigenvectors in Eq.~\eqref{dpp} can be replaced by those based on the original eigenvectors.  

As a proof of concept, this study focuses exclusively on trivial solutions, which are computationally more tractable than hairy solutions. For the Schwarzschild-(A)dS background, $X$ is proportional to the identity matrix and therefore commutes with any matrix. This simplifies  Eq.~\eqref{perX} into
\begin{align}
    \label{dx}
       \delta X    = P'   \delta \bar X   {P'^{ - 1}}.
\end{align}
Crucially, this form eliminates the need for an explicit expression for   $\delta P'   {P'^{ - 1}}$,  thereby greatly reducing the computational complexity. We emphasize, however, that this term does not vanish for hairy backgrounds; its treatment will be addressed in future work.

The simplicity of Eq.~\eqref{dx} naturally motivates the question of whether the Schur decomposition can generalize this method to cases where $g^{-1}\eta$ is non-diagonalizable, with $ \bar{X} $ now denoting an upper-triangular matrix.  Indeed, certain trivial solutions exhibit non-diagonalizable $ g^{-1}\eta $ and serve as valid tests for this extension. We have verified that $ X$ can be computed without obstruction in these cases. However, the associated perturbation analysis is sufficiently involved to warrant a separate treatment and lies beyond the scope of the present paper. Details of this generalized method are provided in Appendix~\ref{schur}.

Turning to the perturbations of trivial solutions, we adopt the potential $V(\varphi)=m^2+ V_4 \varphi^4$ and  $W(\varphi)=0$, where $V_4$ is a constant. The effects of alternative potentials $V$ will be examined in Sec.~\ref{stability}. 
On a Schwarzschild-(A)dS background, the first-order perturbed field equations take the form
\begin{align}
\label{master}
\delta G^\mu_{\nu} &=m^2 \delta X^\mu_{\nu},\\
\label{sca_master}
 \frac{1}{{\sqrt { -g } } }\partial _\mu \left( {\sqrt {-g } g^{\mu\nu}  \partial_\nu \delta \varphi } \right) &=  0.
\end{align}

We now consider metric perturbations and decompose  them in appropriate spherical harmonic bases.  For odd parity,  
\begin{align}
    {\delta g_{AB}} &= 0, \nonumber\\
    {\delta g_{Ab}} &= \sum\limits_{lm} {{H_A^{lm}}\left( {t,r} \right)S_b^{lm}},\nonumber\\
    {\delta g_{ab}} &= \sum\limits_{lm} {{K^{lm}}\left( {t,r} \right)S_{ab}^{lm}},
\end{align}
where $S_b^{lm}$ represents the axial vector harmonics  and $S_{ab}^{lm}$ represents the axial rank-2 tensor harmonics.  We adopt the convention that uppercase indices (such as $A$, $B$)  denote $t$ or $r$, while lowercase indices (such as $a$, $b$)  denote the angular coordinates. 
For even parity, 
\begin{align}
    {\delta g_{AB}} &= \sum\limits_{lm} \begin{pmatrix}
        {h_0^{lm}\left( {t,r} \right)}&{h_1^{lm}\left( {t,r} \right)}\\
{h_1^{lm}\left( {t,r} \right)}&{h_2^{lm}\left( {t,r} \right)}
\end{pmatrix}Y^{lm}, \nonumber\\
    {\delta g_{Ab}} &= \sum\limits_{lm} {{h_A^{lm}}\left( {t,r} \right)Y_b^{lm}},\nonumber\\
    {\delta g_{ab}} &= \sum\limits_{lm} {{k^{lm}}\left( {t,r} \right)Z_{ab}^{lm}} + h_3^{lm}\left( {t,r} \right) \gamma_{ab}Y^{lm},
\end{align}
where  $Y^{lm}$ represents the spherical harmonics,  $Y_{b}^{lm}$  represents  the polar vector harmonics, $Z_{ab}^{lm}$ represents the polar rank-2 tensor harmonics and $\gamma_{ab}$ is the metric of the 2-sphere.

Applying this method, we compute   $ \delta X$ for both the special solutions and the general solutions; the results are listed below, with detailed calculations for the special and general solutions provided in Appendices~\ref{ptb_s} and~\ref{ptb_b}, respectively. 
$ \delta X$ stays mixed-index during the calculation, but it is convenient to lower the contravariant index.  

In the case of special solutions with  $\beta = \alpha^2$, for odd parity 
\begin{align}
g_{A\mu }\delta X^\mu_{B} &=g_{A\mu }\delta X^\mu_{b}=0,\nonumber\\
g_{a\mu }\delta X^\mu_{b}  &=  \sum\limits_{lm} \frac{{{\kappa _{12}}}}{{2f({k_3}-1)}}{K^{lm}S_{ab}^{lm}},
\end{align}
where $\kappa_{ij}$ is defined in Appendix~\ref{ptb_s},   for even parity 
\begin{align}
    g_{A\mu }\delta X^\mu_{B} &= g_{A\mu }\delta X^\mu_{b}=0, \nonumber\\
g_{a\mu }\delta X^\mu_{b}  &=  \sum\limits_{lm} \frac{{{\kappa _{12}}}}{{2f({k_3}-1)}}({k^{lm}Z_{ab}^{lm}}- h_3^{lm}\gamma_{ab}Y^{lm}).
\end{align}
In the case of general solutions,  assuming $k_2=k_3$, for odd parity 
\begin{align}
    g_{A\mu }\delta X^\mu_{B} &=  0, \nonumber\\
    g_{r\mu }\delta X^\mu_{b} &=  \frac{\kappa_{13}}{4 (k_3-1)fx}\sum\limits_{lm} \left((x+y)H_r^{lm}+2 bH_t^{lm}\right)S_b^{lm},\nonumber\\
     g_{t\mu }\delta X^\mu_{b} &= -\frac{2b}{x+y} g_{r\mu }\delta X^\mu_{b},\nonumber\\
     g_{a\mu }\delta X^\mu_{b}  &=  \frac{\kappa_{13}}{2 (k_3-1)f} \sum\limits_{lm} {{K^{lm}}S_{ab}^{lm}}.
\end{align}
For even parity,
\begin{align}
    g_{A\mu }\delta X^\mu_{B} &=  -\frac{\kappa_{13}}{ (k_3-1)f^2r^2}\sum\limits_{lm} \begin{pmatrix}
        \frac{ (y-x)}{2 x}&-\frac{b }{x} \\[1ex]
-\frac{b }{x} &\frac{ (x+y)}{2 x}
\end{pmatrix}h_3^{lm}Y^{lm}, \nonumber\\ 
g_{r\mu }\delta X^\mu_{b} &=  \frac{\kappa_{13}}{4 (k_3-1)fx}\sum\limits_{lm} \left((x+y)h_r^{lm}+2 bh_t^{lm}\right)Y_b^{lm},\nonumber\\
     g_{t\mu }\delta X^\mu_{b} &= - \frac{2b}{x+y}g_{r\mu }\delta X^\mu_{b},\nonumber\\
     g_{a\mu }\delta X^\mu_{b}  &=  \frac{\kappa_{13}}{2 (k_3-1)f} \sum\limits_{lm}\biggl[ {{k^{lm}}Z_{ab}^{lm}} -h_3 ^{lm}\gamma_{ab}Y^{lm} \nonumber\\
    & +\frac{r^2}{2  x}\left( (x-y)h_{0}^{lm}-4 b h_{1}^{lm}- (x+y)h_{2}^{lm} \right)\gamma_{ab}Y^{lm} \biggr]. 
\end{align}

As discussed in Appendix~\ref{bgsol}, the degeneracy of $ k_3 $ is $2 $ for hairy solutions. However, as $ \varphi $ —or equivalently, the graviton mass—approaches a constant, the hairy solutions converge exclusively to the general Schwarzschild-(A)dS solutions, where the degeneracy of $k_3$ increases to $3$. This discontinuity potentially suggests a singular perturbation.  At first glance,  both terms in $\delta X$ appear problematic:  the eigenvectors in   $P'   \delta \bar X   {P'^{ - 1}} $ undergo an abrupt change,  while
$\left[\delta P'   {P'^{ - 1}}, X  \right]$ does not suffer from this issue; however,  certain terms of the form  $\frac{1}{\lambda_i-\lambda_j}$ diverge as the degeneracy of $k_3$ increases. 
The resolution may lie in the possibility that, despite these individual singularities, their combination yields a smooth result.   

 In Appendix~\ref{cgml}, we  prove that both $\delta X$ and $\delta U$ are indeed smooth in the limit. Consequently, we conclude that this apparent singularity is merely a mathematical artifact, and the method remains applicable to scenarios involving this limit, such as spontaneous scalarization.

Further complications arise when the degeneracy of $k_3$ 
reaches  $4$, corresponding to solutions where the eigendecomposition method fails. Nevertheless, we can prove the validity of these solutions using Schur decomposition,  which constitutes a low-dimensional subspace of the solution space. Their perturbations can be analyzed under specific limiting procedures. Details are provided in Appendix~\ref{schur}.

In summary, the trivial solutions encompass all possible degeneracies of $k_3$ ($2$, $3$ and $4$), in spherically symmetric setups, exhibiting rich perturbative structures. On the other hand, despite their higher computational cost, perturbations of hairy solutions are conceptually simpler due to the fixed degeneracy.

\section{Stability of trivial black holes } \label{stability}
In this section, we examine the stability of trivial black holes by drawing direct conclusions from the perturbation equations.  
Eq.~\eqref{master} shows that the metric perturbation decouples completely from the scalar sector; it can therefore be solved independently as a  perturbation equation in GR with a source term arising from the graviton potential.   Obtaining a solution is nontrivial, except for a specific case.
For a general background, the parameter choice  $\beta = \alpha ^2$ directly implies   
$\delta X  = 0$.  As indicated  by Eq.~\eqref{perdx_g}, $\delta X $ is proportional to $\kappa_{13}$, which vanishes under this parameter choice.  Alternatively, in a special background one can always choose  $e(r)$ such that the solution approaches the general branch, as the corresponding constraint vanishes. This yields  $k_2 \rightarrow k_3$ or $k_1 \rightarrow k_3$, both of which enforce  $\kappa_{12} \rightarrow 0$, and consequently   $\delta X  \rightarrow  0$, as indicated  by Eq.~\eqref{44}.  In this sense, the solution manifold of general solutions and that of special solutions intersect only at this specific locus, where  $\delta X  = 0$.  Therefore, the perturbation remains smooth during the transition between general and special solutions in either limit,  since its trajectory necessarily passes through $\delta X  = 0$.
Even if the eigenvalue degeneracies of $\mathcal{K}$ could potentially change again in this limit, further analysis would be unnecessary.

Since the parameter choice  $\beta = \alpha ^2$ does not admit solutions with static scalar hair, as shown in Appendix~\ref{bgsol},  it is therefore expected that the trivial Schwarzschild-(A)dS solutions under this parameter choice are free from instabilities induced by nontrivial background scalar configurations, since there is no static scalar profile to trigger them. However, this absence of hair does not guarantee dynamical stability against perturbations whose effective mass squared becomes negative due to unfavorable scalar potential choices.  In what follows, we examine the stability of these trivial solutions. 

Under the  condition $\delta X=0$, Eq.~\eqref{master} becomes  identical to the metric perturbation equation in GR.  This conclusion seems to be  consistent with \cite{Kodama:2013rea}. Similarly, the scalar perturbation Eq.~\eqref{sca_master} becomes identical to that of a massless minimally coupled scalar field in GR; it must therefore yield identical stability properties. Four-dimensional Schwarzschild-(A)dS black holes are known to be linearly stable under gravitational perturbations. In addition,  trivial solutions of a minimally coupled scalar field on these backgrounds are linearly stable under the appropriate boundary conditions \cite{Kodama:2003jz, Dafermos:2007jd, Holzegel:2011rk}. 

Choosing $V$ as $V(\varphi)=m^2+ \frac{1}{2}V_2 \varphi^2$, the scalar perturbation equation takes the form
\begin{align}
     \frac{1}{{\sqrt { -g } } }\partial _\mu \left( {\sqrt {-g } g^{\mu\nu}  \partial_\nu \delta \varphi } \right) &=  -V_2  U \delta \varphi.  
\end{align}
in which the term $-V_2  U $ acts as the effective mass squared for the scalar perturbation. For solutions with $\beta = \alpha ^2$, $U$ is given by 
\begin{align}
  U=  \frac{1}{\alpha ^2}\left(\frac{\alpha +1}{\sqrt{c_0}}-\alpha \right),
\end{align}
where $c_0$ is defined in Appendix \ref{bgsol}.
On a Schwarzschild-dS background, to avoid tachyonic instabilities for the scalar, the fundamental requirement is that this effective mass squared remains non-negative,
which is satisfied when $\alpha>0$, $c_0>\frac{(\alpha +1)^2}{\alpha ^2}$, $V_2>0$.
On a Schwarzschild-AdS background, stability can be assessed via the Breitenlohner–Freedman (BF) bound. It can be satisfied by multiple parameter regimes, which are lengthy and omitted here.

If we set $m^2=0$, the background reduces to  Minkowski spacetime and  the trivial black holes  disappear.  The only possible black hole must have hair. 

\section{conclusion} \label{conclusion}

In this paper we have developed a new method to calculate the first-order  perturbation in mass-varying massive gravity, applicable to both dRGT theory and its extensions. This approach provides a unified framework for handling  $X$, $U$, $\delta X$, and $\delta U$,  thereby streamlining the perturbation analysis of hairy  solutions. 

For background solutions that satisfy $ X^\mu_\nu \propto \delta ^\mu_\nu$, such as the Schwarzschild-(A)dS background, this method is advantageous   because it automatically eliminates the commutator term  $\left[\delta P'   {P'^{ - 1}}, X  \right]$.  Other methods do not distinguish this special class of backgrounds; consequently,   it is reasonable to assume that perturbation calculations typically retain this commutator despite its vanishing value, introducing unnecessary computational overhead.  Our method avoids this redundancy and remains computationally efficient even when applied to more complex backgrounds. 

Although the diagonalizability of   $\tilde \Omega $   is required to initiate the calculation, the explicit forms of its eigenvectors never enter the subsequent derivation.   Only  
 $\tilde \Omega$ and its trace are needed,  both of which are well-defined provided that  
${\cal K}$ is diagonalizable. We have explicitly verified this condition for degeneracies of  $2$ and $3$. It is straightforward to extend this verification to cases with fourfold degeneracy or two distinct eigenvalues each with twofold degeneracy.  In the absence of degeneracy, the construction of  $\tilde \Omega $ becomes entirely unnecessary.  

A limitation of this method is the requirement that  ${\cal K}$ be diagonalizable, which precludes certain reference metric choices. A notable example arises in bi-gravity black hole studies, where conventional reference metrics render 
${\cal K}$ non-diagonalizable.  Furthermore, the method is not directly applicable to abstract perturbations $h_{\mu\nu}$,  since the closed-form analytic expressions for the   eigenvalues and eigenvectors of ${\cal K}$ are generally unattainable. 
Nevertheless, it may be possible to deduce from  Eq.~\eqref{perX}  a general expression involving only   $g_{\mu\nu}$ and $h_{\mu\nu}$ by eliminating the spectral dependence. Importantly, the computational efficiency of our method stems precisely from its tailored treatment of specific metrics, enabling progressive simplification throughout the derivation rather than requiring global simplification at the final stage.  

As an application, we have analyzed perturbations of trivial Schwarzschild-(A)dS black holes and found that for the special parameter choice  
 $\beta = \alpha ^2$,  the system reduces to GR on a Schwarzschild-(A)dS background, minimally coupled to a scalar field whose mass is determined by the integration constant $c_0$,  the parameter $\alpha$ and the potential parameters. Under appropriate conditions, these black holes are stable. While the perturbation equations for generic parameter values have been derived, obtaining their full solution remains a nontrivial task, which will be explored in upcoming studies.  Equipped with this method, future investigations can systematically address the stability of both trivial and hairy black holes, as well as other scalar-related phenomena in MVMG.  

\section*{Acknowledgement}
We would like to thank Shuang-Yong Zhou for comments.  We thank the Elite Revitalizing Inner Mongolia Program (2025TGL05) for their support.

\appendix

\section*{APPENDICES}

\subsection{Background solutions} \label{bgsol}

The $^t_r$ component of the field equations is 
\begin{align}
\label{fenlei}
{b\left( {\beta {{k_3}}^2  + 2\alpha {{k_3}} + 1} \right)=0}.
\end{align} 
It implies that $
{{k_3}} = \frac{\pm\sqrt{\alpha ^2-\beta }-\alpha }{\beta }
$ in non-diagonal solutions since the possibility of  $b=0$ has been excluded,  and the solutions split into two branches. The conclusion holds regardless of the configuration of the scalar field given it is static. It can be solved directly that
\begin{align}
    f(r) = \frac{1}{(k_3-1)^2}.
\end{align}
To recover the Schwarzschild-(A)dS solution,  we set $\varphi=0$,  which is a trivial solution of Eq.~\eqref{eompsi}.  The solution for $a(r)$  and $c(r)$ can be obtained
\begin{align}
    a\left( r \right)& = {{{c}}_0}{\left( {{{{k}}_3} - 1} \right)^2}- \frac{{M}}{r} + \frac{1}{6}{m^2}(\alpha {{{k}}_3} + 2){{{c}}_0}{{{k}}_3}{r^2},\nonumber\\ 
    c\left( r \right)& =  c_0,
\end{align}
where $c_0$ and $M$ are integration constants.  Furthermore, it can be verified that  
\begin{align}
    G^\mu_\nu= m^2 X^\mu_\nu=\frac{{{m^2}}}{2}{k_3}\left( {\alpha {k_3} + 2} \right)  \delta^\mu_\nu,
\end{align}
where $\delta^\mu_\nu$ is the identity matrix. The $ ^\theta_\theta$ component of the field equations introduces a constraint
\begin{align}
\label{caeq}
    (k_3-k_1)  (k_3-k_2)  (1 + \alpha k_3)=0.
\end{align}
This equation divides the solutions into two categories: one requires that $ k_3=k_1$ or $ k_3=k_2$, the other $ k_3=-\frac{1}{\alpha}$. 

In the first case,  the following solution can be obtained
\begin{align}
e\left( r \right)& =  -a\left( r \right)+ \frac{1}{{{{\left( {{k_3} - 1} \right)}^2}}} + {c_0}{\left( {{k_3} - 1} \right)^2},
\end{align}
where  $c_0$ is a constant of integration. $ k_3=k_1$ or $ k_3=k_2$ gives the same result.  If  we let $ k_2=k_3$ then $k_1=1+\frac{1}{\sqrt{c_0}(k_3-1)}$.  By the definitions of $k_1$ and $k_2$,   $k_1$ is  greater  than $k_2$. If $k_3$ is greater than $1+\frac{1}{\sqrt{c_0}(k_3-1)}$, we simply switch $k_1$ and $k_2$ and the solution remains unchanged. The appearance of two branches of solutions is merely an artifact resulting from assigning $k_1$ and $k_2$ to two different values. There is only one physical solution.  In fact 
\begin{align}
\label{x_para}
    x= \frac{\left |(1-k_3)^4c_0-1 \right |  }{(1-k_3)^2},
\end{align}
where the absolute value originates from the square root. If the absolute value results in a positive sign, then $k_2=k_3$, otherwise $k_1=k_3$.
The solution has one single eigenvalue  and a degenerate eigenvalue of $k_3$ with both algebraic and geometric multiplicity equal to 3.    

If we choose  $ k_3=-\frac{1}{\alpha}$, then $\beta=\alpha^2$ immediately follows, in which case the value of $k_3$ is unique and the two branches of solutions have merged. Also Eq.~\eqref{caeq} vanishes  and there is no constraint left to determine $e(r)$. The simplest choice would be $e(r)=1$,  and the solution  becomes a Schwarzschild-(A)dS solution  in the Gullstrand–Painlevé frame, or to set $e(r)=0$, in which case the solution  becomes  a Schwarzschild-(A)dS solution in the  Eddington-Finkelstein frame.  Consequently  the solution has two single eigenvalues of $k_1$, $k_2$  and a degenerate eigenvalue of $k_3$ with both algebraic and geometric multiplicity equal to 2.   

Regarding the degeneracy of $k_3$, there is another possible case that arises from a solution that does not require a specific choice of $\beta$.  For this reason we categorize the solution as a general one. As a constant of integration, $c_0$ can be chosen such that $k_1 = k_2 = k_3$; upon calculation, we find that $c_0 = \frac{1}{(k_3 - 1)^4}$ satisfies the condition.  The perturbation of this solution should be treated separately because  the degeneracy of $k_3$ is now equal  to $ 4$, which differs from the previous two cases. However, in this solution,  the geometric multiplicity of $k_3$ equals  $3$, and the matrix $\mathcal{K}$  is therefore not diagonalizable.  Other methods such as Schur decomposition should be applied in the calculation instead of  eigendecomposition. We work out some of the details  in Appendix~\ref{schur}. 

Besides the trivial solution $\varphi = 0$, there also exist hairy solutions that can only be obtained by numerical methods.  For hairy solutions, the constraint assumes a more complex form,
\begin{align}
\label{caeq2}
    2(k_3-k_1)  (k_3-k_2)  (1 + \alpha k_3)V+ rk_3(k_3(2+\alpha k_3)-U)\varphi' V'(\varphi)=0.
\end{align}
It  cannot be eliminated  by setting $ k_3=-\frac{1}{\alpha}$, so there is no freedom in choosing the form of $e(r)$.  In this sense, the limit of hairy solutions in which $\varphi$ approaches a constant would be the general solutions, not the special solutions. In fact, substituting $k_3 = -\frac{1}{\alpha}$ into the constraint yields 
\begin{align}
    r \left(\alpha  \sqrt{c}+1\right) \varphi' V'(\varphi)=0,
\end{align}
it can be satisfied by setting $V$ or $c$ to a constant value.  In both cases, the graviton mass is a constant; in the latter case, it is due to one of the field equations $ c'= \frac{1}{2} r c \varphi'^2$  \cite{Tolley:2015ywa}. Hence, there is no hairy solution corresponding to the parameter choice of $\beta=\alpha^2$.  On the other hand, conditions $ k_2=k_3$ or $ k_1=k_3$ directly imply that $V$ is a constant, or that $k_3(2+\alpha k_3)-U=0$, neither of which can be satisfied by a hairy solution. This constraint therefore fixes the degeneracy of $k_3$ in the hairy solutions to 2.

However, when $\varphi$—or equivalently, the graviton mass—approaches a constant, this also leads to the emergence of pathological behavior, namely, the degeneracy of $k_3$ changes from $2$ to $3$. On the surface, this does not appear to be a smooth transition. We will discuss  this issue in Appendix~\ref{cgml}.  
 
\subsection{Perturbations} \label{ptb_a}

In this appendix, we present a detailed calculation of $\delta X^\mu_{\nu}$ for different backgrounds. 
It is convenient to lower the contravariant index of $\delta X^{\mu}_{\nu}$  and express it in terms of spherical harmonics once the calculation is completed.   Because $\delta X_{\mu\nu} =  \delta g_{\mu \lambda} X^\lambda_\nu+  g_{\mu \lambda} \delta X^\lambda_\nu$ and $X^\lambda_\nu $ is proportional to the identity matrix,  $g_{\tau\mu }\delta X^\mu_{\nu} $ is symmetric. Therefore, it should decompose readily into harmonics.  We also set $\delta g_{\mu\nu} = h_{\mu\nu}$ and use $\gamma_{ab}$ to  denote the metric of 2-sphere.  This section deals with matrix operations;  we number the matrix indices from $1$ to $4$ for convenience.  We have maintained the convention that uppercase indices (such as $A$, $B$)  denote $t$ or $r$, while lowercase indices (such as $a$, $b$)  denote the angular coordinates.

\subsubsection{Special solutions } \label{ptb_s}
In this background,  $k_1$ and $k_2$ are different while $ k_3=k_4$. The  non-zero elements of $\delta \bar X$ are
\begin{align}
\label{dx_g_h}
 \delta \bar X_{11} &= \kappa_{23}(\delta k_3+\delta k_4), \nonumber \\
   \delta \bar X_{22} &= \kappa_{13}(\delta k_3+\delta k_4),\nonumber \\
   \delta \bar X_{33} &=\kappa_{12} \delta k_4+\kappa_{13} \delta k_2
   +\kappa_{23} \delta k_1, \nonumber\\
   \delta \bar X_{44} &=\kappa_{12}  \delta k_3 +\kappa_{13}\delta k_2
   +\kappa_{23} \delta k_1,
\end{align}
in which we have defined $ \kappa_{ij}=\frac{1}{2}  (\alpha  (k_i+k_j)+\beta  k_i k_j+1)$.  In the special Schwarzschild-(A)dS background, the values of $k_1$ and $k_2$ are known functions of $r$.  We can use the condition $\beta=\alpha^2$ and $k_3=-\frac{1}{\alpha}$ to obtain a rather simple result
\begin{align}
\label{xdij}
   \delta \bar X_{11} &= 0, \nonumber \\
   \delta \bar X_{22} &=0,\nonumber \\
   \delta \bar X_{33} &=\kappa_{12} \delta k_4, \nonumber\\
   \delta \bar X_{44} &=\kappa_{12}  \delta k_3,
\end{align}
because with such parameter a choice,   $\kappa_{ij}=\frac{1}{2}(\alpha k_i+1)(\alpha k_i+1)$, and it would equal $0$ when $i$ or $j$  is $3$. 
Using the eigenvectors given by \eqref{eigenvt}, $\tilde \Omega $  is defined as the $2 \times 2$ bottom-right block of  $\delta g^{\mu \rho }  \eta_{\rho\nu}$ corresponding to $\theta$ and $\varphi$, 
\begin{align}
    \tilde \Omega = -\frac{1}{f^2 r^2} \gamma^{-1} \begin{pmatrix}
        h_{33} &  h_{34}\\
      h_{43}  &  h_{44}
    \end{pmatrix}= -\frac{1}{f^2 r^2}
    \begin{pmatrix}
        h_{33} &  h_{34}\\
        \csc^2(\theta)h_{43}  &  \csc^2(\theta) h_{44}
    \end{pmatrix}.
\end{align}
The form of $\tilde \Omega$ depends on the choice of eigenvectors, but it can be shown that the final expression for $\delta X$ does not. 

The matrix of $\delta X^\mu_\nu$ should be block diagonal. To demonstrate this, we can rewrite Eq.~\eqref{dx} in block matrix form,
\begin{align}
   \renewcommand{\arraystretch}{1.5}
    \left[
\begin{array}{c|c}
\delta\mathcal{X}_{tr} & 0  \\
\hline
0  & \delta \mathcal{X}_{\theta \varphi}\\
\end{array}
\right]=
\left[
\begin{array}{c|c}
\mathcal{P} _{tr}& 0  \\
\hline
0  & \mathcal{P}'_{\theta \varphi} \\
\end{array}
\right]
\left[
\begin{array}{c|c}
 0 & 0  \\
\hline
0  & \delta \mathcal{\bar X}_{\theta \varphi}\\
\end{array}
\right]
\left[
\begin{array}{c|c}
\mathcal{P}^{-1}_{tr} & 0  \\
\hline
0  & \mathcal{P}'^{-1}_{\theta \varphi}\\
\end{array}
\right].
\end{align}
The non-zero element  represents $2 \times 2$ block of the corresponding matrix, whose position is indicated by its index, while zero element represents $2 \times 2$ zero matrix. $P'$ can be written in this block diagonal form because the last two eigenvectors are mixed only with themselves, while the first two eigenvectors remain unchanged. It is obvious that,
\begin{align}
\label{dxtr}
\delta \mathcal{X}_{tr}=0.
\end{align}
 From Eq.~\eqref{xdij}, 
\begin{align}
\label{xdtp}
 \delta \mathcal{\bar X}_{\theta \varphi} &= \frac{{{\kappa _{12}}}}{{2({k_3}-1)}} {\begin{pmatrix}
    {\delta {\lambda _4}}&0\\
0&{\delta {\lambda _3}}
\end{pmatrix}}\nonumber\\ 
 &=  \frac{{{\kappa _{12}}}}{{2({k_3}-1)}}\left[\left( {\delta {\lambda_3} + \delta {\lambda_4}} \right)I  -  {\begin{pmatrix}
    {\delta {\lambda _3}}&0\\
0&{\delta {\lambda _4}}
\end{pmatrix}} \right],
\end{align}
where $I$ represents $2 \times 2$ identity matrix and we have used the relation  $\delta k_i= \frac{1}{2 (k_i-1)} \delta \lambda_i$.  After multiplication with $\mathcal{P}'_{\theta \varphi}$ and $\mathcal{P}'^{-1}_{\theta \varphi}$, $ I$  remains unchanged, while the second term in the  bracket  becomes $\tilde \Omega$.  This can be verified by observing  that $\mathcal{P}'_{\theta \varphi} $ can be written as  $\mathcal{P}'_{\theta \varphi} = \mathcal{P}_{\theta \varphi} \mathcal{A}$, where $\mathcal{A}$ is a matrix with   eigenvectors of  $\tilde \Omega$ as its column vectors, while $\mathcal{P}_{\theta \varphi}$  simply equals  $I$.  The expression happens to be the eigendecomposition of $\tilde \Omega$.  Factor $ \delta \lambda_3 + \delta \lambda_4$  is  the trace of $\tilde \Omega $ which is $-\gamma^{ab}h_{ab}/f ^2r^2$.  After combining the two terms and lowering the  index, $ \delta \mathcal{X}_{\theta \varphi}$ can be written as
\begin{align}
\label{44}
\frac{{{\kappa _{12}}}}{{2f({k_3}-1)}}\left(h_{ab}-\gamma^{cd}h_{cd}\gamma_{ab}    \right).
\end{align}
In summary, the result is proportional to a trace-reversed $h_{ab}$. 

The vector expansions for both even and odd parity are zero, as indicated by the block diagonal form of  $\delta X^\mu_\nu$. Consequently,  $g_{\tau\mu }\delta X^\mu_{\nu} $  admits a parity-preserving harmonic expansion.     

In the case of odd parity we have 
\begin{align}
    \delta \lambda_1 = \delta \lambda_2 = \delta \lambda_3+\delta \lambda_4=0,
\end{align}
$\delta \lambda_1 = \delta \lambda_2=0$ because the $2 \times 2$ top left block of $\delta g^{-1} \eta$ is zero in odd parity,  $ \delta \lambda_3+\delta \lambda_4=0$ stems from the fact that the trace of tensor expansion is $0$ in odd parity.  
For even parity,
\begin{align}
\label{46}
\delta \lambda_1 &= 2\frac{ (x+y)h_{11}+ 2 b \left(h_{12}+h_{21}\right)+ (y-x)h_{22}}{x (x+z)^2}, \nonumber\\   
\delta \lambda_2 &=2\frac{(x-y) h_{11}- 2 b \left(h_{12}+h_{21}\right) - (x+y)h_{22}}{x (x-z)^2},\nonumber \\ 
    \delta \lambda_3+\delta \lambda_4 &= -\frac{1}{f^2 r^2} \gamma^{ab}h_{ab}.
\end{align}

On a hairy background, $\delta U$ is also  required in the perturbation equation of the scalar. It is given by 
\begin{align}
    \delta U= \frac{1}{\alpha^2}\left[(1+\alpha)\left( (k_1-1) \delta k_2 + (k_2-1) \delta k_1\right)- \alpha(k_1-1)(k_2-1)(\delta k_3 +\delta k_4)  \right].
\end{align}
A notable feature of $\delta \lambda_i$ is that it only contains  scalar harmonics of the metric perturbation. In the odd-parity expansion, all relevant terms are zero.  This is expected since the scalar perturbation should only involve scalar harmonics. This property extends to perturbations in the general solution.   

\subsubsection{General  solutions} \label{ptb_b}
We follow the same method to compute the perturbations of general solutions,  in which  $k_2= k_3=k_4$ while $k_1$ is different.  It implies that $ 2f=z-x$, which will be utilized in later calculations. The non-zero elements of $\delta \bar X$ are 
\begin{align}
\label{perdx_g}
 \delta \bar X_{11} &= 0, \nonumber \\
   \delta \bar X_{22} &=\kappa_{13}(\delta k_3 +\delta k_4),\nonumber \\
   \delta \bar X_{33} &=\kappa_{13}(\delta k_2 +\delta k_4), \nonumber\\
   \delta \bar X_{44} &=\kappa_{13}(\delta k_2 +\delta k_3).
\end{align}
Using the eigenvectors given by \eqref{eigenvt},  the following expression for $\tilde{\Omega}$ can be obtained, 
\begin{align}
\left(
\begin{array}{ccc}
 \dfrac{ (x-y)h_{11}-2 b \left(h_{12}+h_{21}\right)- (x+y)h_{22}}{2 f^2 x} & -\dfrac{2 b h_{13}+ (x+y)h_{23}}{2 f^2 x} & -\dfrac{2 b h_{14}+(x+y)h_{24} }{2 f^2 x} \\[2ex]
 -\dfrac{2 b h_{31}+ (x+y)h_{32}}{f^2 r^2 (x+y)} & -\dfrac{h_{33}}{f^2 r^2} & -\dfrac{h_{34}}{f^2 r^2} \\[2ex]
 -\dfrac{\csc ^2(\theta ) \left(2 b h_{41}+(x+y)h_{42} \right)}{f^2 r^2 (x+y)} & -\dfrac{h_{43} \csc ^2(\theta )}{f^2 r^2} & -\dfrac{h_{44} \csc ^2(\theta )}{f^2 r^2} \\
\end{array}
\right).
\end{align}
We  rewrite \eqref{dx} in a block matrix form 
\begin{align}
\label{dx2}
\delta X
=&
P'
\renewcommand{\arraystretch}{1.5}
\left[
\begin{array}{c|c}
 0 & 0  \\
\hline
0  & \delta \mathcal{\bar X} _{r\theta \varphi}\\
\end{array}
\right]
{P'}^{-1} \nonumber\\ 
=&
\kappa_{13}(\delta k_2 +\delta k_3+\delta k_4)
P'
\renewcommand{\arraystretch}{1.5}
\left[
\begin{array}{c|c}
 0 & 0  \\
\hline
0  & I\\
\end{array}
\right]
{P'}^{-1}
-
\kappa_{13}P'
\left[
\begin{array}{c|c}
 0 & 0  \\
\hline
0  & \delta \mathcal{\bar K}_{r\theta \varphi}\\
\end{array}
\right]
{P'}^{-1},
\end{align}
where $I$ is a $3 \times 3$  identity matrix, $\delta \mathcal{\bar X} _{r\theta \varphi}= \mathrm{diag}( \delta \bar X_{22},  \delta \bar X_{33},  \delta \bar X_{44} )$  is a   $3 \times 3$ diagonal matrix, $  \delta \mathcal{\bar K}_{r\theta \varphi}=\mathrm{diag}( \delta k_2, \delta k_3, \delta k_4)$ is also  a   $3 \times 3$ diagonal matrix. Factor $(\delta k_2 + \delta k_3 + \delta k_4)$ is given by
\begin{align}
\label{51}
\delta k_2 + \delta k_3 + \delta k_4 &= 
 \frac{1}{2 (k_3-1)}   [\tilde{\Omega}] \nonumber \\
 &=   \frac{1}{2 (k_3-1)} \left(\frac{ (x-y)h_{11}-2 b \left(h_{12}+h_{21}\right)- (x+y)h_{22}}{2 f^2 x} -\frac{1}{f^2 r^2}\gamma^{ab}h_{ab}\right),
\end{align}
this term would vanish for odd parity.

The first term in the \eqref{dx2} can be expanded as follows, omitting the factor $\kappa_{13}(\delta k_2 + \delta k_3 + \delta k_4)$, 
\begin{align}
\label{xfirst}
    P'
\renewcommand{\arraystretch}{1.5}
\left[
\begin{array}{c|c}
 0 & 0  \\
\hline
0  & I\\
\end{array}
\right]
{P'}^{-1}= {P}
\renewcommand{\arraystretch}{1.5}\left[
\begin{array}{c|c}
 1 & 0  \\
\hline
0  & \mathcal {A}\\
\end{array}
\right]
\left[
\begin{array}{c|c}
 0 & 0  \\
\hline
0  & I\\
\end{array}
\right]
\left[
\begin{array}{c|c}
 1 & 0  \\
\hline
0  &  \mathcal{A}^{-1}\\
\end{array}
\right]
{P}^{-1}
=   P
\renewcommand{\arraystretch}{1.5}
\left[
\begin{array}{c|c}
 0 & 0  \\
\hline
0  & I\\
\end{array}
\right]
P^{-1},
\end{align}
in which we have used the relations  $P'=PA$ and  $I$ is a $3 \times 3$  identity matrix. The matrix $A$ can be written in the block diagonal form  because the last three eigenvectors mixed up while the first one remain unchanged.  After lowering the index, the result is a 
matrix we denote as $X^{\rm I}$,
\begin{align}
\label{X1}
X^{\rm I}=\left(
\begin{array}{cccc}
 \dfrac{ (y-x)f}{2 x} & -\dfrac{b f}{x} & 0 & 0 \\[2ex]
 -\dfrac{b f}{x} & \dfrac{ (x+y)f}{2 x} & 0 & 0 \\[2ex]
 0 & 0 & f r^2 & 0 \\
 0 & 0 & 0 & f r^2 \sin ^2(\theta ) \\
\end{array}
\right).
\end{align}
The second term can be simplified in a similar approach, 
\begin{align}
    P'
\renewcommand{\arraystretch}{1.5}
\left[
\begin{array}{c|c}
 0 & 0  \\
\hline
0  & \delta \mathcal{\bar K}_{r\theta \varphi}\\
\end{array}
\right]{P'}^{-1}  =&  P
\renewcommand{\arraystretch}{1.5}
\left[
\begin{array}{c|c}
 1 & 0  \\
\hline
0  &  \mathcal{A}\\
\end{array}
\right]
\left[
\begin{array}{c|c}
 0 & 0  \\
\hline
0  & \delta \mathcal{\bar K}_{r\theta \varphi}\\
\end{array}
\right]\left[
\begin{array}{c|c}
 1 & 0  \\
\hline
0  & \mathcal{A}^{-1}\\
\end{array}
\right]
P^{-1}\nonumber\\ 
=& \frac{1}{2 (k_3-1)}P
\renewcommand{\arraystretch}{1.5}
\left[
\begin{array}{c|c}
 0 & 0  \\
\hline
0  & \tilde{\Omega}\\
\end{array}
\right]P^{-1}.
\end{align}
In the second line, we have used the definition of $\delta \lambda_i$ and its relationship with $\delta k_i$. 
After lowering the index we obtain a symmetric matrix $\frac{1}{2 (k_3-1)}X^{\rm II}$, with
\begin{align}
\label{X2}
X^{\rm II}=-\left(
\begin{array}{cccc}
 X^{\rm II}_{11} & X^{\rm II}_{12} & -\dfrac{2b}{x+y} X^{\rm II}_{23} & -\dfrac{2b}{x+y} X^{\rm II}_{24} \\[2ex]
* & X^{\rm II}_{22} & \dfrac{ (x+y)h_{23}+2 b h_{13}}{2 f x} & \dfrac{ (x+y)h_{24}+2 b h_{14}}{2 f x} \\[2ex]
 * &* & \frac{1}{f}h_{33} & \frac{1}{f} h_{34}\\
 *&* & \frac{1}{f}h_{43} & \frac{1}{f}h_{44} \\
\end{array}
\right),
\end{align}
in which
\begin{align}
X^{\rm II}_{11} &=(x-y)\frac{ (x-y) h_{11}-2 b \left(h_{12}+h_{21}\right)- (x+y)h_{22}}{4 f x^2},\nonumber\\
    X^{\rm II}_{12} &=\frac{2 b}{x-y}X^{\rm II}_{11},\nonumber\\ 
    X^{\rm II}_{22} &= \frac{x+y}{y-x}X^{\rm II}_{11}.
\end{align}
The final form of $\delta X$ with lower  indices  is a symmetric matrix and can be written in the following block form
\begin{align}
\label{57}
    \renewcommand{\arraystretch}{1.5}
 \frac{\kappa_{13}}{2 (k_3-1)}\left[
\begin{array}{c|c}
  -\frac{1}{f^2 r^2}\gamma^{ab}h_{ab}X^{\rm I}_{AB} &- X^{\rm II}_{Ab}  \\
\hline
*  &  \frac{1}{f}h_{ab}+[\tilde{\Omega}]fr^2\gamma_{ab}\\
\end{array}
\right].
\end{align}
The right bottom block  can be further reduced to 
\begin{align}
    \frac{1}{f}(h_{ab}-\gamma^{cd}h_{cd}\gamma_{ab} )+r^2 \frac{ (x-y)h_{11}-2 b \left(h_{12}+h_{21}\right)- (x+y)h_{22}}{2 f x}\gamma_{ab}.
\end{align}
The first term can be interpreted as a trace-reversed $h_{ab}$, the second term is a pure trace term.  A notable result is that $ X^{\rm II}_{AB}$ has been canceled out during the calculation.

$\delta k_1$ is the same as in the special solution and $\delta k_2+  \delta k_3+ \delta k_4$ is given by \eqref{51}.
The expression of $\delta U$ is
\begin{align}
\label{59}
   \delta U= -k_3 ((\alpha -1) k_3+2) ( (k_3-1) \delta k_1 +(k_1-1) (\delta k_2+\delta k_3+\delta k_4)).
\end{align}

The $\delta X$ of solutions with $k_1=k_3$ is exactly the same as  \eqref{57}.  It can be readily obtained   by exchanging $k_1$, $v_1$, $u_1$ and  $k_2$, $v_2$, $u_2$.  Further analysis reveals that this exchange is equivalent to the transformation  $x\rightarrow-x$. 
As shown in Eqs.~\eqref{k1} and \eqref{k2}, reversing the sign of $x$ interchanges  $k_1$ and $k_2$. This transformation also exchanges the eigenvectors; taking  $v_1$ as an example, 
\begin{align}
    \left(1,\frac{2b}{x+y},0,0\right)^T
    =\frac{2b}{x+y}\left(\frac{x+y}{2b},1,0,0\right)^T
    =\frac{2b}{x+y}\left(\frac{2b}{y-x},1,0,0\right)^T,
\end{align}
in which the identity $y^2-x^2=4b^2$ has been used. It shows that under the  transformation,    $v_1$ becomes proportional to   $v_2$. The proportionality factor is physically irrelevant and can be omitted.  The left eigenvectors transform analogously.  Overall, while  \eqref{57} is  calculated under the assumption that $k_2=k_3$, to obtain the expression of  $\delta X$ corresponding to $k_1=k_3$, one only needs to make a transformation  $x\rightarrow-x$. However, as can be seen from  Eq.~\eqref{x_para},  $x$ contains an absolute value,  which yields a sign  change when switching from $k_2=k_3$ to $k_1=k_3$, canceling the transformation, leaving the final expression invariant. The same conclusion applies to   $\delta  U$.  

However, for the sake of simplicity, if  $x$ is retained in the expression—as we did in previous sections—then the perturbations of $k_1=k_3$ differ in the sign of $x$.

\subsubsection{The constant graviton mass limit } \label{cgml}

As demonstrated in Appendix~\ref{bgsol}, in the limit of constant mass, $k_2$ or $k_1$ approaches $k_3$ and the degeneracy of $k_3$ grows from $2$ to $3$.  In this section, we will examine whether this limit is smooth in the current  method.

Consider a hairy solution  $g'_{\mu\nu}$  decomposed as 
\begin{align}
    g'_{\mu\nu}= g_{\mu\nu}+  h^*_{\mu\nu},
\end{align}
where  $ g_{\mu\nu}$ is a trivial solution. Suppose $g'_{\mu\nu}$ also satisfies
\begin{align}
k_2&= k_3+\delta,\nonumber\\
\lambda_2&=\lambda_3 + 2(k_2-1)\delta.
\end{align}
Both $h^*_{\mu\nu}$ and $\delta$ are small. To study $\delta X$ in the constant mass limit,  we introduce a regular perturbation  $h_{\mu\nu}$ on top of the hairy background  and use Eq.~\eqref{perX} to expand $X$ in terms of $h_{\mu\nu}$, treating  $k_2$ and $k_3$ as  distinct eigenvalues. 
We then take the limit as $\delta \rightarrow 0$, setting $ h^*_{\mu\nu}=0$. In this limit,   all the terms should be smooth except  those involving  factors such as $ \frac{1}{\lambda_2-\lambda_3}$. For smooth terms, the limit is  evaluated by  setting  $k_2=k_3$ and $\lambda_2=\lambda _3$; singular terms, however, require careful analysis.  

In the limit,  $ P'   \delta \bar X   {P'^{ - 1}}$ in Eq.~\eqref{perX} is smooth. 
$\delta \bar X$ is given by setting $k_2=k_3$ in  Eq.~\eqref{dx_g_h}, 
\begin{align}
 \delta \bar X_{11} &= 0, \nonumber \\
   \delta \bar X_{22} &=\kappa_{13}(\delta k_3 +\delta k_4),\nonumber \\
   \delta \bar X_{33} &=\kappa_{13}(\delta k_2 +\delta k_4), \nonumber\\
   \delta \bar X_{44} &=\kappa_{13}(\delta k_2 +\delta k_3).
\end{align}
It is identical to Eq.~\eqref{perdx_g}, however $\delta k_i$ is defined differently. Because $k_2$ and $k_3$ are treated as  distinct eigenvalues, the degeneracy of $k_3$ remains $2$, thus $ \tilde \Omega $ is a $2\times2$ matrix given by
\begin{align}
    \tilde \Omega = -\frac{1}{f^2 r^2} \gamma^{-1} \begin{pmatrix}
        h_{33} &  h_{34}\\
      h_{43}  &  h_{44}
    \end{pmatrix},
\end{align}
from which we can obtain 
\begin{align}
    \delta k_3+ \delta k_4= \frac{1}{2 (k_3-1)} ( \delta \lambda_3+ \delta \lambda_4)=-\frac{1}{2 (k_3-1)}\frac{1}{f^2 r^2} \gamma^{ab}h_{ab}.
\end{align}
$\delta X$ can be written in a block  matrix form
\begin{align}
   \renewcommand{\arraystretch}{1.5}
    \left[
\begin{array}{c|c}
\delta\mathcal{X}_{tr} & 0  \\
\hline
0  & \delta \mathcal{X}_{\theta \varphi}\\
\end{array}
\right]=
\left[
\begin{array}{c|c}
\mathcal{P} _{tr}& 0  \\
\hline
0  & \mathcal{P}'_{\theta \varphi} \\
\end{array}
\right]
\left[
\begin{array}{c|c}
 \delta  \mathcal{\bar X}_{tr} & 0  \\
\hline
0  & \delta \mathcal{\bar X}_{\theta \varphi}\\
\end{array}
\right]
\left[
\begin{array}{c|c}
\mathcal{P}^{-1}_{tr} & 0  \\
\hline
0  & \mathcal{P}'^{-1}_{\theta \varphi}\\
\end{array}
\right].
\end{align}
As in Appendix~\ref{ptb_s},  the element represents a 2 × 2 block of the corresponding matrix, whose position is
indicated by its index.  
Direct calculation shows that,  after lowering the index, $\delta \mathcal{X}_{tr}$ is given by
\begin{align}
\kappa_{13} X^{\rm I}_{AB}(\delta k_3+ \delta k_4)=-\frac{\kappa_{13}}{2 (k_3-1)}\frac{1}{f^2 r^2} \gamma^{ab}h_{ab}X^{\rm I}_{AB},
\end{align}
where $X^{\rm I}_{AB}$ is defined in \eqref{X1}. 

 $\delta \mathcal{\bar X}_{\theta \varphi}$ can be written as 
\begin{align}
{{\kappa _{13}}}\left[ \frac{1}{{2({k_3}-1)}}{\begin{pmatrix}
    {\delta {\lambda _4}}&0\\
0&{\delta {\lambda _3}}
\end{pmatrix}}+\frac{1}{{2({k_2}-1)}}\delta {\lambda _2} I\right],
\end{align}
where $I$ is a $2 \times 2$ identity matrix.  After multiplication with  $ \mathcal{P}'_{\theta \varphi}$, $\mathcal{P}'^{-1}_{\theta \varphi}$ and lowering the index, the first term becomes 
\begin{align}
\frac{{{\kappa _{13}}}}{{2f({k_3}-1)}}\left(h_{ab}-\gamma^{cd}h_{cd}\gamma_{ab}    \right),
\end{align}
which is almost identical to Eq.~\eqref{44}   since the calculation is exactly the same, except that $ k_2=k_3$ is used in the constant graviton mass limit.  After the same procedure, the second term becomes
\begin{align}
\frac{{{\kappa _{13}}}}{{2({k_3}-1)}}fr^2\delta \lambda_2\gamma_{ab},
\end{align}
in which $\delta \lambda_2$ is given by Eq.~\eqref{46}.  It can be easily verified that the results  already match the $tr$ and $\theta \varphi$ blocks of $\delta X$ given by \eqref{57}.

We proceed to examine  $\left[\delta P'   {P'^{ - 1}}, X  \right]$ which contains terms that diverge in the limit, due to the presence of $\lambda_2-\lambda_3$ or $\lambda_2-\lambda_4$ in the denominator.
However, further analysis reveals that the result should be finite in the limit.
Specifically, the divergent part  in  $\delta P' {{P'}^{-1}}$ scales as $\mathcal{O}(\frac{1}{\delta})$, while $X \sim I+ \mathcal{O}(\delta)$ because the solution approaches Schwarzschild-(A)dS in this limit.  Consequently, as $\delta$ approaches $0$,  the leading $\mathcal{O}(1)$ part of $X$ commutes with $\delta P' {{P'}^{-1}}$.  The commutator therefore receives contributions only from the $\mathcal{O}(\delta)$ correction to $X$,  canceling the $\mathcal{O}(\frac{1}{\delta})$ divergence  of $\delta P' {{P'}^{-1}}$.
In summary, 
\begin{align}
    \left[\delta P'   {P'^{ - 1}}, X  \right] \sim \left[\mathcal{O}(\delta^{-1}),\mathcal{O}(\delta)  \right] +\mathcal{O}(\delta)\sim\mathcal{O}(1),
\end{align}
 and the result is manifestly finite. 
 
 To determine its expression, we identify the divergent terms in $\delta P'   {P'^{ - 1}}$, these are
\begin{align}
    \frac{{\rm P}_2   \delta {g^{ - 1}}\eta    {\rm P}_3}{{\lambda _3} - {\lambda _2}}+ \frac{{\rm P}_3   \delta {g^{ - 1}}\eta    {\rm P}_2}{{\lambda _2} - {\lambda _3}}+
    \frac{{\rm P}_2   \delta {g^{ - 1}}\eta    {\rm P}_4}{{\lambda _4} - {\lambda _2}}+ \frac{{\rm P}_4   \delta {g^{ - 1}}\eta    {\rm P}_2}{{\lambda _2} - {\lambda _4}},
\end{align}
in which ${\rm P}_i $ is calculated using the eigenvectors in  \eqref{eigenvt}.   We also isolate the  terms in $X$ that are in the order of $\delta$ and disregard the remaining terms. Finally we take the commutator and lower the index,  resulting in the following  symmetric matrix,
\begin{align}
    \renewcommand{\arraystretch}{1.5}
 \frac{\kappa_{13}}{2 (k_3-1)}\left[
\begin{array}{c|c}
 0& -X^{\rm II}_{Ab}  \\
\hline
*  &  0\\
\end{array}
\right],
\end{align}
where $X^{\rm II}_{AB}$ is defined in \eqref{X2}. The result happens to be the vector portion of $\delta X$.

In summary,  starting from a hairy solution, we recover the $\delta X$ of a general Schwarzschild-(A)dS background, confirming that the transition is smooth at least at leading order.   This demonstrates that the present method can handle changes in eigenvalue degeneracy.  

The smoothness of  $\delta U$ remains to be verified, and can be readily established.  We begin with a general form of $\delta U$ on a hairy background and  take the limit by substituting   $k_2 = k_3 = k_4$. The result coincides with   Eq.~\eqref{59},  except that  $\delta k_i$ is initially  calculated based on the condition that $k_2 \neq k_3$. 
We now show that this yields the same result as the  case of  $k_2 = k_3$. In the case of $k_2 \neq k_3$,
\begin{align}
    \delta k_1&=\frac{1}{2(k_1-1)}u_1 \delta g^{-1 }  \eta  v_1,\nonumber\\
    \delta k_2&=\frac{1}{2(k_2-1)}u_2 \delta g^{-1 }  \eta  v_2\nonumber,\\
    \delta k_2+  \delta k_3&=  \frac{1}{2(k_3-1)} \sum\limits_{i=3}^{4}u_i \delta g^{-1 }  \eta  v_i.
\end{align}
whereas for  $k_2 = k_3$.
\begin{align}
    \delta k_1&=\frac{1}{2(k_1-1)}u_1 \delta g^{-1 }  \eta  v_1,\nonumber\\
\delta k_2+  \delta k_3+ \delta k_4&= \frac{1}{2(k_3-1)} \sum\limits_{i=2}^{4} u_i \delta g^{-1 }  \eta  v_i.
\end{align}
Essentially, these two cases  yield identical $\delta U$; the only difference lies in how the components of  $\delta U$ are mapped into different eigenvalue perturbations.  

\subsection{Schur decomposition} \label{schur}

In this appendix we discuss  a peculiar solution with  $c_0 = \frac{1}{(k_3 - 1)^4}$, whose existence cannot be confirmed by  utilizing eigendecomposition.  We try to circumvent the problem by using Schur decomposition.

 $c_0 = \frac{1}{(k_3 - 1)^4}$ is equivalent to  $b= \pm \frac{a-e}{2}$, which yields two identical eigenvalues of $\mathcal{K}$, given by 
\begin{align}
    k_1=k_2= 1- \sqrt{\frac{2}{z}}.
\end{align}
However, there is only one non-zero eigenvector corresponding to the eigenvalues, namely $v_1=(1,1,0,0)^T$. 
The decomposition of $\mathcal{K}$  can be achieved in the following way. The matrix is still in a block diagonal form, and the $\theta \varphi$ block is still diagonalizable.  We only need to apply Schur decomposition to the $tr$ block, or alternatively, to the entire matrix—the result is the same. The Schur form of the  $tr$  block can be written as
\begin{align}
    p\begin{pmatrix} k_1 & k_{12}\\ 0 &k_2 \end{pmatrix}p^{-1},
\end{align}
where $p$ is given by $\frac{1}{\sqrt2}\begin{pmatrix} 1 & -1\\ 1 & 1 \end{pmatrix}$. Using the definition of $\mathcal{K}$,  it can be shown  that 
\begin{align}
    k_{12}= \sqrt{2} (a-e) \left(\frac{1}{z}\right)^{3/2}=2\sqrt{2}b \left(\frac{1}{z}\right)^{3/2},
\end{align}
where we have assumed $b=  \frac{a-e}{2}$.
 $\mathcal{K}$ can be written as 
\begin{align}
    \mathcal{K}=P
    \begin{pmatrix}
        k_1&  k_{12}& 0& 0&\\
         0&  k_2& 0& 0&\\
          0&  0& k_3& 0&\\
           0&  0& 0& k_4&\\
    \end{pmatrix}P^{-1}.
\end{align}
$P$ is given by 
\begin{align}\frac{1}{\sqrt{2}}
    \begin{pmatrix}
        1&  -1& 0& 0&\\
         1& 1& 0& 0&\\
          0&  0& 1& 0&\\
           0&  0& 0& 1&\\
    \end{pmatrix}.
\end{align}
It can be utilized to compute $X$,  yielding field equations identical to those in  Appendix~\ref{bgsol},  which have the same  Schwarzschild-(A)dS  solutions,  where $c_0$ is no longer an integration constant. Thus the solution is valid.  

For Eq.~\eqref{perX} to be valid in the context of  Schur decomposition,
$\delta P'$  should be small. However, we have not yet found a way to ensure this condition. What we can show is that if the perturbations stay within the solution manifold, then $\delta P'$ is small.  Furthermore,   it is necessary to determine the eigenvalue perturbation of a defective matrix,  which is a topic  beyond the scope of this paper. Consequently, even if the Schur decomposition were theoretically applicable to computing   $\delta X$, it  would remain impractical.

These solutions only have one integration  constant, while the general solutions have two; they  essentially reside on a  "boundary" between $k_1=k_3$ and $k_2=k_3$ regimes. Assuming smoothness at this boundary, we could evaluate perturbations by taking an appropriate limit.  The boundary can be reached by substituting $c_0 = \frac{1}{(k_3 - 1)^4}$ into Eq.~\eqref{x_para}, yielding $x = 0$.  Since    \eqref{57} contains several terms with $x$ in the denominator,  it must be evaluated in a limit $x \rightarrow 0$.

Let $x =  \delta$ be a small parameter.    To maintain   $k_2 = k_3$, we simultaneously set  $z = z_0 + \delta$, where $z_0=2f$, which is solved from $k_1=k_2=k_3$ at  $x=0$.
In the limit $x \rightarrow 0$,  the  factor $\kappa _{13}$ in \eqref{57}  is given by
\begin{align}
\label{k13_sch}
  \kappa _{13} =  -\frac{  (\alpha +\beta  k_3)}{\sqrt{2}z^{3/2}}\delta,
\end{align}
The matrix part of \eqref{57}  contains terms of order  $\mathcal{O}( \delta^{-1})$.  When these terms are combined with $\kappa _{13}$, a finite result was obtained, which can be read directly from   \eqref{57}  and is therefore omitted here. One notable fact is that $\delta X$ still vanishes for $\beta = \alpha^2$ due to Eq.~\eqref{k13_sch}.  

In the limit, Eq.~\eqref{59} reduces to
\begin{align}
    \delta U= -\frac{1}{2} k_3 ((\alpha -1) k_3+2) (\delta \lambda_1+\delta \lambda_2+\delta\lambda_3+\delta \lambda_4),
\end{align}
where $\sum \limits_{i } \delta \lambda_i \equiv [\delta g^{-1}\eta]$ is  finite and well-defined.  
The limit is smooth even though individually  $\delta \lambda _1$ and $\delta \lambda _2+\delta \lambda _3+\delta \lambda _4$ all diverge, as can be seen from Eqs.~\eqref{46} and \eqref{51}.  The reason is that
the divergent contributions from  $u_1 \delta g^{-1 }  \eta  v_1$ and $u_2 \delta g^{-1 }  \eta  v_2$ cancel exactly.  In fact,  their sum is identical to  $ (\delta g^{-1}\eta)^1_1 + (\delta g^{-1}\eta)^2_2$,  which is manifestly finite and independent of the limiting procedure.  

In summary, the validity of this solution can be verified by Schur decomposition; its perturbation can be obtained in the limit $x\rightarrow 0$.


\begin{thebibliography}{99}
\bibitem{LIGOScientific:2016aoc}
B.~P.~Abbott \textit{et al.} [LIGO Scientific and Virgo],
Phys. Rev. Lett. \textbf{116}, no.6, 061102 (2016)
doi:10.1103/PhysRevLett.116.061102
[arXiv:1602.03837 [gr-qc]].

\bibitem{Cardoso:2025npr}
V.~Cardoso, S.~Biswas and S.~Sarkar,
[arXiv:2511.14841 [gr-qc]].



\bibitem{LIGOScientific:2016lio}
B.~P.~Abbott \textit{et al.} [LIGO Scientific and Virgo],
Phys. Rev. Lett. \textbf{116}, no.22, 221101 (2016)
[erratum: Phys. Rev. Lett. \textbf{121}, no.12, 129902 (2018)]
doi:10.1103/PhysRevLett.116.221101
[arXiv:1602.03841 [gr-qc]].

\bibitem{LIGOScientific:2021sio}
R.~Abbott \textit{et al.} [LIGO Scientific, VIRGO and KAGRA],
Phys. Rev. D \textbf{112}, no.8, 084080 (2025)
doi:10.1103/PhysRevD.112.084080
[arXiv:2112.06861 [gr-qc]].

\bibitem{Yunes:2024lzm}
N.~Yunes, X.~Siemens and K.~Yagi,
[arXiv:2408.05240 [gr-qc]].

\bibitem{deRham:2010ik}
  C.~de Rham, G.~Gabadadze,
  Phys.\ Rev.\  {\bf D82}, 044020 (2010),
  [arXiv:1007.0443].

\bibitem{deRham:2010kj}
  C.~de Rham, G.~Gabadadze and A.~J.~Tolley,
  Phys.\ Rev.\ Lett.\  {\bf 106}, 231101 (2011),
  [arXiv:1011.1232].

\bibitem{Hassan:2011hr}
  S.~F.~Hassan and R.~A.~Rosen,
  Phys.\ Rev.\ Lett.\  {\bf 108}, 041101 (2012)
  [arXiv:1106.3344 [hep-th]].

\bibitem{Hassan:2011ea}
  S.~F.~Hassan and R.~A.~Rosen,
  JHEP {\bf 1204} (2012) 123
  [arXiv:1111.2070 [hep-th]].

\bibitem{Hinterbichler:2011tt}
K.~Hinterbichler,
Rev. Mod. Phys. \textbf{84}, 671-710 (2012)
doi:10.1103/RevModPhys.84.671
[arXiv:1105.3735 [hep-th]].

\bibitem{deRham:2014zqa}
C.~de Rham,
Living Rev. Rel. \textbf{17}, 7 (2014)
doi:10.12942/lrr-2014-7
[arXiv:1401.4173 [hep-th]].

\bibitem{Babichev:2015xha}
E.~Babichev and R.~Brito,
Class. Quant. Grav. \textbf{32}, 154001 (2015)
doi:10.1088/0264-9381/32/15/154001
[arXiv:1503.07529 [gr-qc]].

\bibitem{Wood:2024acv}
K.~Wood, P.~M.~Saffin and A.~Avgoustidis,
Phys. Rev. D \textbf{109}, no.12, 124006 (2024)
doi:10.1103/PhysRevD.109.124006
[arXiv:2402.17835 [gr-qc]].

\bibitem{Guarato:2013gba}
P.~Guarato and R.~Durrer,
Phys. Rev. D \textbf{89}, no.8, 084016 (2014)
doi:10.1103/PhysRevD.89.084016
[arXiv:1309.2245 [gr-qc]].

\bibitem{Kodama:2013rea}
H.~Kodama and I.~Arraut,
PTEP \textbf{2014}, 023E02 (2014)
doi:10.1093/ptep/ptu016
[arXiv:1312.0370 [hep-th]].

\bibitem{Bernard:2014bfa}
L.~Bernard, C.~Deffayet and M.~von Strauss,
Phys. Rev. D \textbf{91}, no.10, 104013 (2015)
doi:10.1103/PhysRevD.91.104013
[arXiv:1410.8302 [hep-th]].

\bibitem{Bernard:2015mkk}
L.~Bernard, C.~Deffayet and M.~von Strauss,
JCAP \textbf{06}, 038 (2015)
doi:10.1088/1475-7516/2015/06/038
[arXiv:1504.04382 [hep-th]].

\bibitem{Kobayashi:2015yda}
T.~Kobayashi, M.~Siino, M.~Yamaguchi and D.~Yoshida,
PTEP \textbf{2016}, no.10, 103E02 (2016)
doi:10.1093/ptep/ptw145
[arXiv:1509.02096 [gr-qc]].

\bibitem{Cusin:2015tmf}
G.~Cusin, R.~Durrer, P.~Guarato and M.~Motta,
JCAP \textbf{04}, 051 (2016)
doi:10.1088/1475-7516/2016/04/051
[arXiv:1512.02131 [astro-ph.CO]].

\bibitem{Bernard:2015uic}
L.~Bernard, C.~Deffayet, A.~Schmidt-May and M.~von Strauss,
Phys. Rev. D \textbf{93}, no.8, 084020 (2016)
doi:10.1103/PhysRevD.93.084020
[arXiv:1512.03620 [hep-th]].

\bibitem{Mazuet:2017hey}
C.~Mazuet and M.~S.~Volkov,
Phys. Rev. D \textbf{96}, no.12, 124023 (2017)
doi:10.1103/PhysRevD.96.124023
[arXiv:1708.03554 [hep-th]].


\bibitem{Mazuet:2018ysa}
C.~Mazuet and M.~S.~Volkov,
JCAP \textbf{07}, 012 (2018)
doi:10.1088/1475-7516/2018/07/012
[arXiv:1804.01970 [hep-th]].

\bibitem{Wood:2025ujj}
K.~Wood,
Phys. Rev. D \textbf{113}, no.8, 084054 (2026)
doi:10.1103/vspq-pkfy
[arXiv:2509.15055 [hep-th]].

\bibitem{Huang:2012pe}
Q.~G.~Huang, Y.~S.~Piao and S.~Y.~Zhou,
Phys. Rev. D \textbf{86}, 124014 (2012)
doi:10.1103/PhysRevD.86.124014
[arXiv:1206.5678 [hep-th]].

\bibitem{Tolley:2015ywa}
A.~J.~Tolley, D.~J.~Wu and S.~Y.~Zhou,
Phys. Rev. D \textbf{92}, no.12, 124063 (2015)
doi:10.1103/PhysRevD.92.124063
[arXiv:1510.05208 [hep-th]].


\bibitem{Zhang:2017jze}
J.~Zhang and S.~Y.~Zhou,
Phys. Rev. D \textbf{97}, no.8, 081501 (2018)
doi:10.1103/PhysRevD.97.081501
[arXiv:1709.07503 [gr-qc]].

\bibitem{Koyama:2011yg}
K.~Koyama, G.~Niz and G.~Tasinato,
Phys. Rev. D \textbf{84}, 064033 (2011)
doi:10.1103/PhysRevD.84.064033
[arXiv:1104.2143 [hep-th]].

\bibitem{Berezhiani:2011mt}
L.~Berezhiani, G.~Chkareuli, C.~de Rham, G.~Gabadadze and A.~J.~Tolley,
Phys. Rev. D \textbf{85}, 044024 (2012)
doi:10.1103/PhysRevD.85.044024
[arXiv:1111.3613 [hep-th]].

\bibitem{Kodama:2003jz}
H.~Kodama and A.~Ishibashi,
Prog. Theor. Phys. \textbf{110}, 701-722 (2003)
doi:10.1143/PTP.110.701
[arXiv:hep-th/0305147 [hep-th]].

\bibitem{Dafermos:2007jd}
M.~Dafermos and I.~Rodnianski,
[arXiv:0709.2766 [gr-qc]].

\bibitem{Holzegel:2011rk}
G.~Holzegel and J.~Smulevici,
Commun. Math. Phys. \textbf{317}, 205-251 (2013)
doi:10.1007/s00220-012-1572-2
[arXiv:1103.3672 [gr-qc]].


\end{thebibliography}
\end{document}